\documentclass[aps,prd,reprint,superscriptaddress]{revtex4-2}

\usepackage{amsmath}
\usepackage{graphicx}
\usepackage{dcolumn}
\usepackage{siunitx}
\usepackage{fancyvrb}
\usepackage{mdframed}
\usepackage[colorlinks=true,allcolors=blue,pdfpagelabels=false]{hyperref}

\newcommand{\mc}[1]{\multicolumn{1}{c}{#1}}
\newcolumntype{d}[1]{D{+}{\,\pm\,}{#1}}

\begin{document}

\title{Transverse charge density and the radius of the proton}

\author{Alexander V. Gramolin}
\affiliation{Department of Physics, Boston University, Boston, Massachusetts 02215, USA}

\author{Rebecca L. Russell}
\affiliation{The Charles Stark Draper Laboratory, Inc., Cambridge, Massachusetts 02139, USA}

\begin{abstract}
A puzzling discrepancy exists between the values of the proton charge radius obtained using different experimental techniques: elastic electron-proton scattering and spectroscopy of electronic and muonic hydrogen. The proton radius is defined through the slope of the electric form factor, $G_E(Q^2)$, at zero four-momentum transfer, which is inaccessible in scattering experiments. We propose a novel method for extracting the proton radius from scattering data over a broad $Q^2$ range rather than attempting to directly determine the slope of $G_E$ at $Q^2 = 0$. This method relates the radius of the proton to its transverse charge density, which is the two-dimensional Fourier transform of the Dirac form factor, $F_1(Q^2)$. We apply our method to reanalyze the extensive data obtained by the A1 Collaboration [J.~C.~Bernauer \textit{et~al.}, \href{https://doi.org/10.1103/PhysRevLett.105.242001}{Phys. Rev. Lett. \textbf{105}, 242001 (2010)}] and extract a radius value, $r_E = 0.889(5)_{\text{stat}}(5)_{\text{syst}}(4)_{\text{model}}~\text{fm}$, that is consistent with the original result. We also provide new parametrizations for the Dirac and Pauli form factors and the transverse charge and magnetization densities of the proton. Our reanalysis shows that the proton radius discrepancy cannot be explained by issues with fitting and extrapolating the A1 data to $Q^2 = 0$.
\end{abstract}

\maketitle

\section{Introduction}

Over a century after Rutherford's discovery of the proton~\cite{Rutherford_PhilosMag.37.581}, some fundamental properties of this particle are still not well understood. In particular, the proton charge radius, $r_E$, remains experimentally puzzling. Beginning with the pioneering research~\cite{Hofstadter_PR.98.217, Hofstadter_RMP.28.214}, $r_E$ has long been measured in elastic electron-proton scattering experiments~\cite{Simon_NuclPhysA.333.381, Bernauer_PRL.105.242001, Bernauer_PRC.90.015206, Zhan_PLB.705.59, Mihovilovic_PLB.771.194, Mihovilovic_EPJA.57.107, Xiong_Nature.575.147}. It has also been extracted from atomic transition frequencies in both electronic~\cite{Mohr_RMP.88.035009, Beyer_Science.358.79, Fleurbaey_PRL.120.183001, Bezginov_Science.365.1007, Grinin_Science.370.1061} and muonic~\cite{Pohl_Nature.466.213, Antognini_Science.339.417} hydrogen. The 2014 CODATA recommended value of $r_E$, obtained from all nonmuonic data available at the time, is $0.8751(61)~\text{fm}$~\cite{Mohr_RMP.88.035009}. In contrast, muonic hydrogen spectroscopy yielded the value $r_E = 0.84087(39)~\text{fm}$~\cite{Antognini_Science.339.417}, which is smaller by 5.6 standard deviations. More recently, there have been experimental results in favor of both the smaller~\cite{Beyer_Science.358.79, Bezginov_Science.365.1007, Xiong_Nature.575.147, Grinin_Science.370.1061} and larger~\cite{Fleurbaey_PRL.120.183001, Mihovilovic_EPJA.57.107} values of the proton radius. The striking discrepancy between different measurements of $r_E$ has become known as the ``proton radius puzzle''~\cite{Pohl_ARNPS.63.175, Carlson_PPNP.82.59, Karr_NatRevPhys.2.601, Gao_RMP.94.015002}. In this paper, we propose a novel method for extracting $r_E$ from scattering data and use it to reanalyze the measurement reported in Refs.~\cite{Bernauer_PRL.105.242001, Bernauer_PRC.90.015206}.

The electromagnetic structure of the proton is encoded in its Dirac and Pauli form factors, $F_1(Q^2)$ and $F_2(Q^2)$, which depend on the negative four-momentum transfer squared, $Q^2 = -q^2$ (see textbooks~\cite{Berestetskii, Halzen&Martin, Thomas&Weise}). Instead of $F_1$ and $F_2$, it is often more convenient to use the Sachs electric and magnetic form factors, defined as
\begin{equation}
G_E = F_1 - \frac{Q^2}{4M^2} \kappa F_2, \qquad G_M = F_1 + \kappa F_2, \label{eq:GE_GM}
\end{equation}
where $M \approx 0.938~\text{GeV}$ is the mass of the proton and $\kappa \approx 1.793$ is its anomalous magnetic moment. The Sachs form factors have a simple interpretation when considered in the Breit frame, where the exchanged virtual photon carries momentum~$\mathbf{q}$ but no energy~\cite{Berestetskii, Halzen&Martin, Thomas&Weise}. In this frame, $Q^2 = \mathbf{q}^2$ and $G_E$ and $G_M$ can be interpreted as the three-dimensional Fourier transforms of the proton's spatial charge and magnetization densities, respectively.

Unfortunately, the concept of the three-dimensional densities is valid only in the nonrelativistic limit, when $Q^2 \ll M^2$ and the Breit frame coincides with the proton rest frame~\cite{Berestetskii, Miller_PRC.99.035202, Vanderhaeghen_NPN.21.14}. For this reason, the proton radius cannot be properly determined through the three-dimensional charge density and is instead defined as
\begin{equation}
r_E = \sqrt{-6 \left. \frac{d G_E(Q^2)}{d Q^2} \right|_{Q^2 = 0}} \label{eq:r_E}
\end{equation}
in both scattering and spectroscopic measurements~\cite{Miller_PRC.99.035202}. However, the definition~(\ref{eq:r_E}) is inconvenient for scattering experiments: it requires measuring $G_E$ at the lowest achievable $Q^2$ values, extrapolating the data down to $Q^2 = 0$, and then inferring the slope of $G_E$ at that point. Such a procedure is inevitably model dependent, which greatly complicates the extraction of the proton radius~\cite{Kraus_PRC.90.045206, Horbatsch_PRC.93.015204, Sick_Atoms.6.2, Yan_PRC.98.025204, Hagelstein_PLB.797.134825, Pacetti_EPJA.56.74}. We propose to avoid these issues by relating the radius of the proton to its transverse charge density, which has a proper relativistic interpretation and can be determined from scattering data over a broad $Q^2$ range.

\section{Transverse charge density}

In this section, we briefly review the definition of the transverse charge density and its relation to the proton radius~\cite{Gao_RMP.94.015002, Miller_PRC.99.035202, Bouchiat_NuclPhysB.34.157, Burkardt_PRD.62.071503, Burkardt_IJMPA.18.173, Miller_PRL.99.112001, Carlson_PRL.100.032004, Miller_ARNPS.60.1, Venkat_PRC.83.015203, Lorce_PRL.125.232002}. We start with a change of space-time coordinates from the usual $(x^0, \, x^1, \, x^2, \, x^3)$ to $(x^{+}, \, x^{-}, \, \mathbf{b})$, where $x^{\pm} = (x^0 \pm x^3) / \sqrt{2}$ are the light-cone variables and $\mathbf{b} = (x^1, \, x^2)$ is the transverse position vector. By setting $q^+ = 0$, we specify the infinite-momentum frame in which $q^{\mu}$ has only transverse components: $q^{\mu} = (0, \, 0, \, \mathbf{q}_{\perp})$ and $Q^2 = \mathbf{q}_{\perp}^2$. Then, the Dirac form factor $F_1(Q^2)$ can be related to a circularly symmetric transverse charge density of the proton, $\rho_1(b)$, by the following two-dimensional Fourier transforms:
\begin{gather}
F_1(Q^2) = 2\pi \int\limits_{0}^{\infty} b \rho_1(b) J_0(Q b) \, db, \label{eq:F1_transform} \\
\rho_1(b) = \frac{1}{2\pi} \int\limits_{0}^{\infty} Q F_1(Q^2) J_0(Q b) \, dQ, \label{eq:rho1_transform}
\end{gather}
where $b = |\mathbf{b}|$ is the impact parameter and $J_0$ denotes the Bessel function of the first kind of order zero. As the forward and inverse Fourier transforms, Eqs.~(\ref{eq:F1_transform}) and (\ref{eq:rho1_transform}) are dual representations of the same quantity in momentum and position spaces. For example, the $m$-pole form factor,
\begin{equation}
F_{m\text{-pole}}(Q^2) = \left(1 + \frac{Q^2}{\Lambda^2}\right)^{-m}, \label{eq:F_multipole}
\end{equation}
which is a generalization of the monopole ($m = 1$) and dipole ($m = 2$) form factors, corresponds to
\begin{equation}
\rho_{m\text{-pole}}(b) = \frac{\Lambda^{m+1} b^{m-1}}{2^m (m-1)! \, \pi} K_{m-1}(\Lambda b),
\end{equation}
where $\Lambda$ is a scale parameter and $K_{m-1}$ denotes the modified Bessel function of the second kind of order $m-1$. Note that $\rho_1(b)$ has a proper density interpretation even in the relativistic case~\cite{Miller_PRC.99.035202}. As a reduction of the generalized parton distributions, it can be related to observables in deep inelastic scattering~\cite{Burkardt_IJMPA.18.173}.

Expanding $J_0(Q b)$, we can rewrite Eq.~(\ref{eq:F1_transform}) as
\begin{equation}
F_1(Q^2) = 1 - \frac{\langle b_1^2 \rangle}{4} Q^2 + \frac{\langle b_1^4 \rangle}{64} Q^4 - \ldots, \label{eq:moment_expansion}
\end{equation}
where
\begin{equation}
\langle b_1^n \rangle = 2\pi \int\limits_0^{\infty} b^{n+1} \rho_1(b) \, db \label{eq:b1_integral}
\end{equation}
is the $n$th moment of $\rho_1(b)$ and $F_1(0) = \langle b_1^0 \rangle = 1$. The moment expansion~(\ref{eq:moment_expansion}) indicates that the mean-square transverse charge radius of the proton is
\begin{equation}
\langle b_1^2 \rangle = -4 \left. \frac{d F_1(Q^2)}{d Q^2} \right|_{Q^2 = 0}. \label{eq:b1_derivative}
\end{equation}
Equation~(\ref{eq:b1_derivative}) is analogous to the proton radius definition~(\ref{eq:r_E}). Note that Eqs.~(\ref{eq:b1_integral}) and (\ref{eq:b1_derivative}) are not equivalent: any fit of the experimental data obtained at $Q^2 > 0$ predicts some slope for $F_1$ at $Q^2 = 0$ but does not necessarily correspond to a bounded transverse charge density. For example, the Fourier integral~(\ref{eq:rho1_transform}) diverges if $F_1$ is a polynomial in~$Q^2$. Therefore, Eqs.~(\ref{eq:rho1_transform}) and (\ref{eq:b1_integral}) impose additional physical constraints on the fit that is used to extract~$\langle b_1^2 \rangle$ (see Appendix~\ref{app:parametrization_choice} for further discussion).

It is important to recognize that there is a simple connection between $r_E$ and $\langle b_1^2 \rangle$. Indeed, after differentiating Eq.~(\ref{eq:GE_GM}) for $G_E$ with respect to $Q^2$, setting $Q^2 = 0$, and substituting Eqs.~(\ref{eq:r_E}) and (\ref{eq:b1_derivative}), we obtain 
\begin{equation}
r_E = \sqrt{\frac{3}{2} \left(\langle b_1^2 \rangle + \frac{\kappa}{M^2}\right)}. \label{eq:r_E_new}
\end{equation}
This equation defines the proton radius through the second moment~(\ref{eq:b1_integral}) of the transverse charge density. Note that we use Eq.~(\ref{eq:b1_derivative}) only to derive the relation~(\ref{eq:r_E_new}) but not to experimentally determine $\langle b_1^2 \rangle$.

\section{Parametrizations for $\rho_1(b)$ and $F_1(Q^2)$}

In principle, one can use Eq.~(\ref{eq:rho1_transform}) to determine $\rho_1(b)$ directly from the experimental data for $F_1(Q^2)$, then calculate $\langle b_1^2 \rangle$ according to Eq.~(\ref{eq:b1_integral}), and finally obtain the proton radius using Eq.~(\ref{eq:r_E_new}). In practice, it is easier to parametrize $F_1(Q^2)$ and $\rho_1(b)$ such that both the Fourier transforms (\ref{eq:F1_transform}) and (\ref{eq:rho1_transform}), as well as the moments~(\ref{eq:b1_integral}), can be calculated analytically. Since $F_1(Q^2)$ at small $Q^2$ is close to the dipole form factor, we expect that $\rho_1(b)$ can be approximated as $\rho_{2\text{-pole}}(b)$ times a polynomial in $\Lambda b$. Particularly suitable are the orthogonal polynomials $P_n^{(\nu)}$ defined by the orthonormality condition
\begin{equation}
\int\limits_0^{\infty} P_m^{(\nu)}(x) \, P_n^{(\nu)}(x) \, w_{\nu}(x) \, dx = \delta_{m n}, \label{eq:orthonormality}
\end{equation}
where
\begin{equation}
w_{\nu}(x) = \frac{2}{\Gamma (\nu + 1)} \, x^{\nu/2} K_{\nu}(2\sqrt{x}) 
\end{equation}
is the weight function, $m$ and $n$ are the degrees of the polynomials, $\delta_{m n}$ is the Kronecker delta, and $\Gamma$ denotes the gamma function. These polynomials have recently been studied in Ref.~\cite{Yakubovich} (note that we use a different normalization for the weight function). We choose $\nu = 1$ and $x = \Lambda^2 b^2 / 4$ to match $w_{\nu}(x)$ with $\rho_{2\text{-pole}}(b)$. The first three corresponding polynomials are
\begin{gather}
P_0^{(1)}(x) = 1, \qquad P_1^{(1)}(x) = \frac{x - 2}{2\sqrt{2}}, \\
P_2^{(1)}(x) = \frac{x^2 - 15 \, x + 18}{6\sqrt{26}}.
\end{gather}
For more terms, see the Supplemental Material~\cite{Supplement}.

We can therefore approximate the transverse charge density as a truncated series
\begin{equation}
\rho_1(b) \approx \rho_{\text{2-pole}}(b) \sum\limits_{n=0}^N \alpha_n P_n^{(1)} \left(\Lambda^2 b^2 / 4\right), \label{eq:rho1_expansion}
\end{equation}
where $\alpha_n$ are the expansion coefficients. After substituting Eq.~(\ref{eq:rho1_expansion}) into Eq.~(\ref{eq:b1_integral}), we find
\begin{equation}
\langle b_1^0 \rangle = \alpha_0, \qquad \langle b_1^2 \rangle = \frac{8}{\Lambda^2} \bigl(\alpha_0 + \sqrt{2} \alpha_1\bigr). \label{eq:b1_moments}
\end{equation}
In general, $\langle b_{1}^{2n} \rangle$ is a linear combination of $\alpha_0, \alpha_1, \dots, \alpha_n$. Therefore, if $\Lambda$ is fixed, there is a one-to-one correspondence between the expansion coefficients~$\alpha_n$ and the even moments of the transverse charge density~(\ref{eq:rho1_expansion}).

After substituting the series expansion~(\ref{eq:rho1_expansion}) into Eq.~(\ref{eq:F1_transform}), we obtain the following parametrization for the Dirac form factor:
\begin{equation}
F_1(Q^2) \approx \sum\limits_{n=0}^N \alpha_n A_n \left(Q^2 / \Lambda^2\right), \label{eq:F1_expansion}
\end{equation}
where
\begin{align}
A_0(y) &= \frac{1}{\left(1 + y\right)^2}, \\
A_1(y) &= -\frac{y \left(y + 4\right)}{\sqrt{2} \left(1 + y\right)^4}, \\
A_2(y) &= \frac{y^2 \left(3 y^2 + 22 y + 39\right)}{\sqrt{26} \left(1 + y\right)^6}, \: \ldots, \\
A_N(y) &= \int\limits_{0}^{\infty} P_{N}^{(1)}(x) \, w_1(x) \, J_0(2\sqrt{x y}) \, dx \label{eq:A_N}
\end{align}
are rational functions.

Extending our formalism to the Pauli form factor, we can represent it as
\begin{equation}
F_2(Q^2) = 2\pi \int\limits_0^{\infty} b \rho_2(b) J_0(Q b) \, db, \label{eq:F2_transform}
\end{equation}
where $\rho_2(b)$ is the transverse magnetization density~\cite{Miller_PRL.101.082002}. It is argued in Ref.~\cite{Miller_ARNPS.60.1} that $\rho_M = -b \left(d\rho_2 / db\right)$ is a better defined quantity. (There is also a closely related transverse charge density of a polarized proton~\cite{Vanderhaeghen_NPN.21.14, Carlson_PRL.100.032004, Gao_RMP.94.015002}.) However, we are not concerned here with the physical interpretation of $\rho_2(b)$ and use it only to parametrize $F_2(Q^2)$.

For reasons that will become clear shortly, we approximate $\rho_2(b)$ as another truncated series,
\begin{equation}
\rho_2(b) \approx \rho_{\text{3-pole}}(b) \sum\limits_{n=0}^N \beta_n P_n^{(2)} \left(\Lambda^2 b^2 / 4\right), \label{eq:rho2_expansion}
\end{equation}
where $\beta_n$ are the expansion coefficients. After substituting this into Eq.~(\ref{eq:F2_transform}), we get
\begin{equation}
F_2(Q^2) \approx \sum\limits_{n=0}^N \beta_n B_n \left(Q^2 / \Lambda^2\right), \label{eq:F2_expansion}
\end{equation}
where
\begin{align}
B_0(y) &= \frac{1}{\left(1 + y\right)^3}, \\
B_1(y) &= -\frac{\sqrt{3} \, y \left(y + 5\right)}{\sqrt{5} \left(1 + y\right)^5}, \\
B_2(y) &= \frac{y^2 \left(7 y^2 + 64 y + 132\right)}{\sqrt{110} \left(1 + y\right)^7}, \: \ldots, \\
B_N(y) &= \int\limits_{0}^{\infty} P_{N}^{(2)}(x) \, w_2(x) \, J_0(2\sqrt{x y}) \, dx. \label{eq:B_N}
\end{align}

We set $\alpha_0 = \beta_0 = 1$ to ensure that $F_1(0) = F_2(0) = 1$. At high~$Q^2$, our parametrizations have the asymptotic behavior expected from the dimensional scaling laws~\cite{Brodsky_PRD.11.1309}: $F_1 \propto (\Lambda / Q)^4$ and $F_2 \propto (\Lambda / Q)^6$. This justifies our choice of the series expansions for $\rho_1(b)$ and $\rho_2(b)$ (other possible parametrizations are discussed in Appendix~\ref{app:parametrization_choice}). Note that the terms $A_0$ and $B_0$ correspond to the dipole and ``tripole'' ($m = 3$) form factors~(\ref{eq:F_multipole}). As shown in the next section and Appendix~\ref{app:consistency}, our parametrizations for $F_1$ and $F_2$ are flexible and efficiently fit experimental data.

\section{Extraction of the proton radius}

Based on the above results, we propose the following method for determining the proton charge radius. First, the measured cross sections are fit with the Rosenbluth formula~(\ref{eq:Rosenbluth}) assuming the parametrizations (\ref{eq:F1_expansion}) and (\ref{eq:F2_expansion}) for the Dirac and Pauli form factors, where $\Lambda$, $\alpha_1, \ldots, \alpha_N$, and $\beta_1, \ldots, \beta_N$ are $2N + 1$ free parameters. Then the mean-square transverse charge radius $\langle b_1^2 \rangle$ is calculated from $\Lambda$ and $\alpha_1$ using Eq.~(\ref{eq:b1_moments}). Finally, the proton radius is given by Eq.~(\ref{eq:r_E_new}). Note that our method does not require measuring the form factor slope at $Q^2 = 0$ and considers the data at all $Q^2$ values. It also allows one to extract the transverse densities $\rho_1(b)$ and $\rho_2(b)$ given by Eqs.~(\ref{eq:rho1_expansion}) and (\ref{eq:rho2_expansion}).

To illustrate our method, we apply it to the extensive and precise elastic electron-proton scattering data obtained by the A1 Collaboration at the Mainz Microtron MAMI~\cite{Bernauer_PRL.105.242001, Bernauer_PRC.90.015206, Bernauer_thesis}. The collaboration measured 1422 cross sections at $Q^2$ values spanning the range from 0.004 to $1~\text{GeV}^2$. Three magnetic spectrometers and six beam energies (180, 315, 450, 585, 720, and 855~MeV) were used, resulting in 18 distinct experimental data groups. To overcome the problem of achieving the absolute normalization of the measurement with subpercent accuracy, they exploited the large redundancy of the data and introduced 31 free normalization parameters, fit simultaneously with the different form factor models. The A1 Collaboration obtained the following value for the proton charge radius:
\begin{equation}
r_E = 0.879(5)_{\text{stat}}(4)_{\text{syst}}(2)_{\text{model}}(4)_{\text{group}}~\text{fm}, \label{eq:A1_radius}
\end{equation}
where the numbers in parentheses represent the statistical, systematic, model, and ``group'' uncertainties. The statistical uncertainty accounts for all point-to-point errors of the cross sections, not only those due to counting statistics. The ``group'' uncertainty was introduced because of an unexplained difference between the radii obtained using the spline and the polynomial groups of form factor models.

\begin{table}
\caption{\label{tab:cross_val}Group-wise cross-validation results for different expansion orders before ($\lambda = 0$) and after ($\lambda > 0$) regularization was applied.}
\begin{ruledtabular}
\begin{tabular}{cccccc}
& \multicolumn{2}{c}{$\lambda = 0$} & \multicolumn{3}{c}{$\lambda > 0$} \\
\cline{2-3} \cline{4-6}
$N$ & $\chi_{\text{train}}^2$ & $\chi_{\text{test}}^2$ & $\lambda$ & $\chi_{\text{train}}^2$ & $\chi_{\text{test}}^2$ \\[2pt]
\hline
1 & 4934 & 5114 & & & \\
2 & 1949 & 2029 & & & \\
3 & 1876 & 2358 & & & \\
4 & 1854 & 2255 & & & \\
5 & 1574 & 1682 & 0.02 & 1574 & 1657 \\
6 & 1566 & 1703 & 0.07 & 1571 & 1664 \\
7 & 1557 & 1912 & 0.2 & 1570 & 1672 \\
8 & 1544 & 2060 & 0.4 & 1569 & 1679 \\
\end{tabular}
\end{ruledtabular}
\end{table}

Following the original analysis~\cite{Bernauer_PRL.105.242001, Bernauer_PRC.90.015206, Bernauer_thesis}, we fit the data with our parametrizations by minimizing the objective function
\begin{equation}
\chi^2 = \sum\limits_{i} \frac{\left(p_i \sigma_{i}^{\text{exp}} - \sigma_{i}^{\text{fit}}\right)^2}{\left(p_i \Delta \sigma_i\right)^2}, \label{eq:chi2}
\end{equation}
where $\sigma_{i}^{\text{exp}}$ are the measured cross sections, $\Delta \sigma_i$ are their point-to-point uncertainties, $\sigma_{i}^{\text{fit}}$ are the model cross sections, and $p_i$ are known combinations of 31 free normalization parameters. The total number of fit parameters is $2N + 32$, where $N$ is the order of the form factor expansions (\ref{eq:F1_expansion}) and (\ref{eq:F2_expansion}). When choosing the value of~$N$, it is important to avoid both underfitting and overfitting---a problem known as the bias-variance trade-off. The popular reduced chi-square test is not appropriate for this purpose because the number of degrees of freedom is ill-defined for a nonlinear fit~\cite{Andrae_arXiv:1012.3754}. Instead, we use cross-validation and regularization, which are standard techniques in statistical learning~\cite{LFD, ISL}.

Careful cross-validation is critical to finding the right balance in the bias-variance trade-off. Typically, a model is cross-validated by randomly dividing the data into $k$ subsets, fitting the model to $k - 1$ of them, testing it on the remaining subset, and repeating the last two steps $k$ times so that each of the subsets is used as a test set exactly once. However, this procedure assumes that errors on the data points are uncorrelated. We instead perform cross-validation by holding out each of the 18 experimental data groups in turn and testing on that group while training on the others. Recall that the data groups correspond to different spectrometer and beam energy combinations. This 18-fold group cross-validation allows us to minimize overfitting to systematic artifacts by ensuring that the model generalizes well to unseen experimental conditions.

The cross-validation results for different orders~$N$ are shown in Table~\ref{tab:cross_val}, where $\chi_{\text{train}}^2$ and $\chi_{\text{test}}^2$ are the total chi-square values~(\ref{eq:chi2}) obtained on the training and test sets, respectively. Note that each data point occurs only once in test sets but 17 times in training sets. For this reason, $\chi_{\text{train}}^2$ has been divided by 17 to make it directly comparable to~$\chi_{\text{test}}^2$. While $\chi_{\text{train}}^2$ monotonically decreases as $N$ increases and the model becomes more flexible, $\chi_{\text{test}}^2$ reaches a minimum at $N = 5$. This indicates underfitting for $N < 5$ and overfitting for $N > 5$.

To control the overfitting in the higher-order models ($N \ge 5$), we add Tikhonov regularization to our objective function:
\begin{equation}
L = \chi^2 + \lambda \sum\limits_{n=1}^{N} \left(\alpha_n^2 + \beta_n^2\right), \label{eq:L}
\end{equation}
where $\alpha_n$ and $\beta_n$ are the expansion coefficients and $\lambda$ is the regularization parameter. The second term in Eq.~(\ref{eq:L}) encourages the sum of the squares of the expansion coefficients to be small and thus reduces the flexibility of the model in a controlled way. We determine the optimal regularization parameter for each order by scanning a range of $\lambda$ values and choosing the one that results in the lowest $\chi_{\text{test}}^2$. One can see from Table~\ref{tab:cross_val} that regularization improves $\chi_{\text{test}}^2$ without significantly compromising $\chi_{\text{train}}^2$. As expected, the optimal $\lambda$ value and the improvement in~$\chi_{\text{test}}^2$ increase with~$N$.

\begin{table}
\caption{\label{tab:results}Objective function values and extracted radii for the regularized models trained on the full dataset.}
\begin{ruledtabular}
\begin{tabular}{cccccc}
$N$ & $\lambda$ & $L$ & $\chi^2$ & $\langle b_1^2 \rangle$ & $r_E$ \\
& & & & $\left(\text{GeV}^{-2}\right)$ & $(\text{fm})$ \\[2pt]
\hline
5 & 0.02 & 1584 & 1576 & 11.49 & 0.889 \\
6 & 0.07 & 1580 & 1573 & 11.42 & 0.887 \\
7 & 0.2  & 1579 & 1572 & 11.37 & 0.885 \\
8 & 0.4  & 1578 & 1571 & 11.32 & 0.883 \\
\end{tabular}
\end{ruledtabular}
\end{table}

\begin{figure*}
\includegraphics[width=\textwidth]{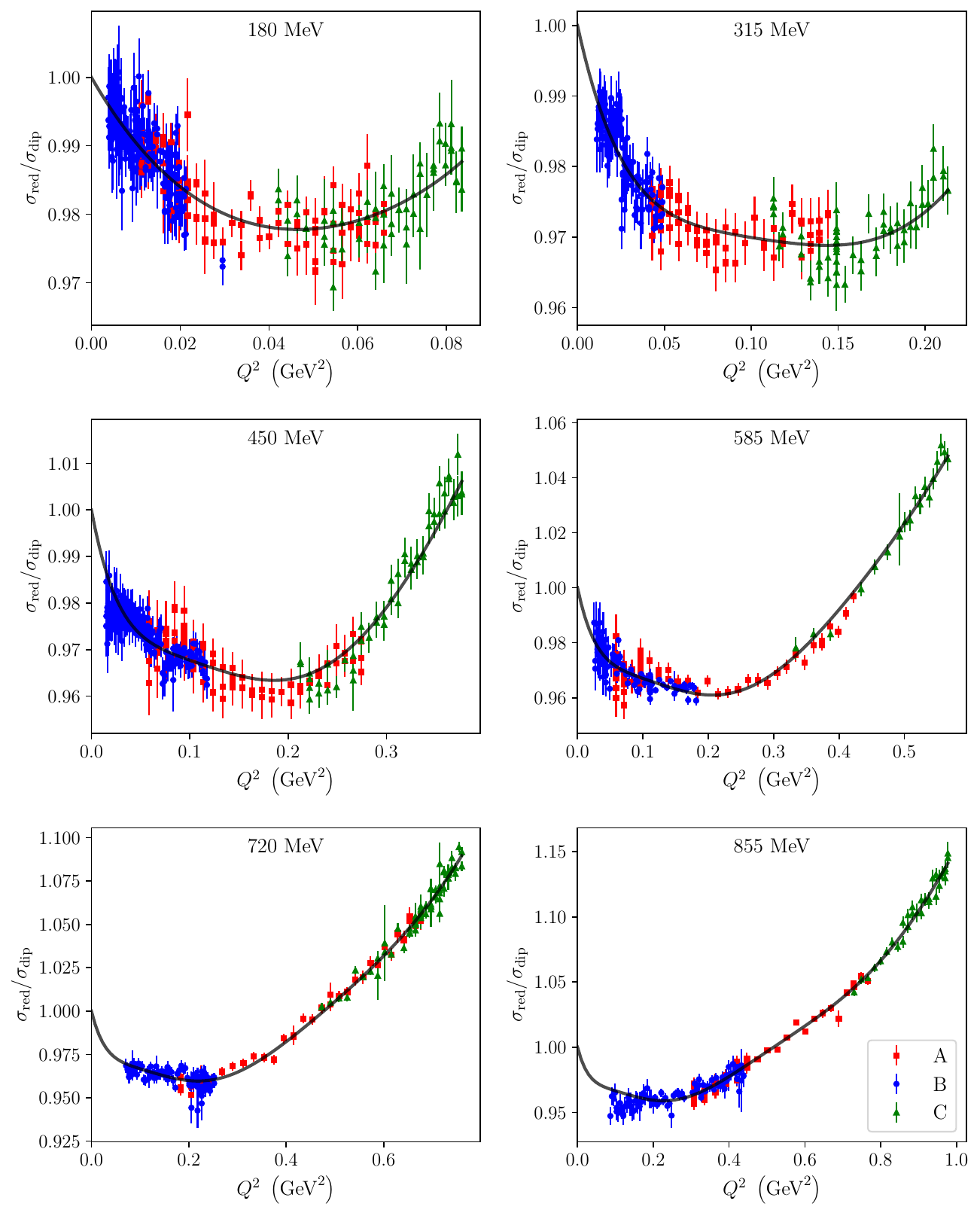}
\caption{\label{fig:cs_fits}Experimental data~\cite{Bernauer_PRC.90.015206} and our best fit as functions of~$Q^2$ for six different beam energies (180, 315, 450, 585, 720, and 855~MeV). The data points are the measured reduced cross sections scaled by our best-fit normalizations and divided by the corresponding dipole cross sections (see Appendix~\ref{app:cross_section}). The different markers represent the spectrometers A, B, and~C. The error bars indicate point-to-point uncertainties of the cross section values.}
\end{figure*}

After the optimal values of $\lambda$ are determined, we train the $N \ge 5$ models on the full dataset (see Table~\ref{tab:results}). We use the $N = 5$ model as our main fit and the higher orders to estimate model misspecification uncertainty (see Appendix~\ref{app:model_dependence} for discussion). Our best fit ($N = 5$, $\lambda = 0.02$) is shown in Fig.~\ref{fig:cs_fits} in comparison with the analyzed cross sections, and the corresponding best-fit parameters are provided in Table~\ref{tab:coeffs}. Note that we achieve a similar $\chi^2$ value to that of Ref.~\cite{Bernauer_PRC.90.015206} while using a more efficient parametrization of the form factors (1576 for 11 parameters vs.\ 1565 for 16 parameters of the spline model). Our cross section normalizations differ from those determined in the original analysis by less than 0.3\%.

\begin{table*}
\caption{\label{tab:coeffs}Expansion coefficients for our best fit. The scale parameter was found to be $\Lambda = 1.156 \pm 0.029~\text{GeV}$.}
\begin{ruledtabular}
\begin{tabular}{c@{\qquad}cd{8,5}d{5}d{5}d{5}d{4}}
& $n = 0$
& \mc{~~~~$n = 1$}
& \mc{$n = 2$}
& \mc{$n = 3$}
& \mc{$n = 4$}
& \mc{~$n = 5$} \\
\hline
$\alpha_n$ & 1 & 0.649 + 0.055 & 1.85 + 0.14 & 6.09 + 0.58 & 9.82 + 1.1 & 5.88 + 0.77 \\
$\beta_n$ & 1 & -0.046 + 0.068 & -2.78 + 0.23 & -8.73 + 0.62 & -11.5 + 0.8 & -5.75 + 0.42 \\
\end{tabular}
\end{ruledtabular}
\end{table*}

\begin{figure*}
\vspace{\baselineskip}
\includegraphics[width=\textwidth]{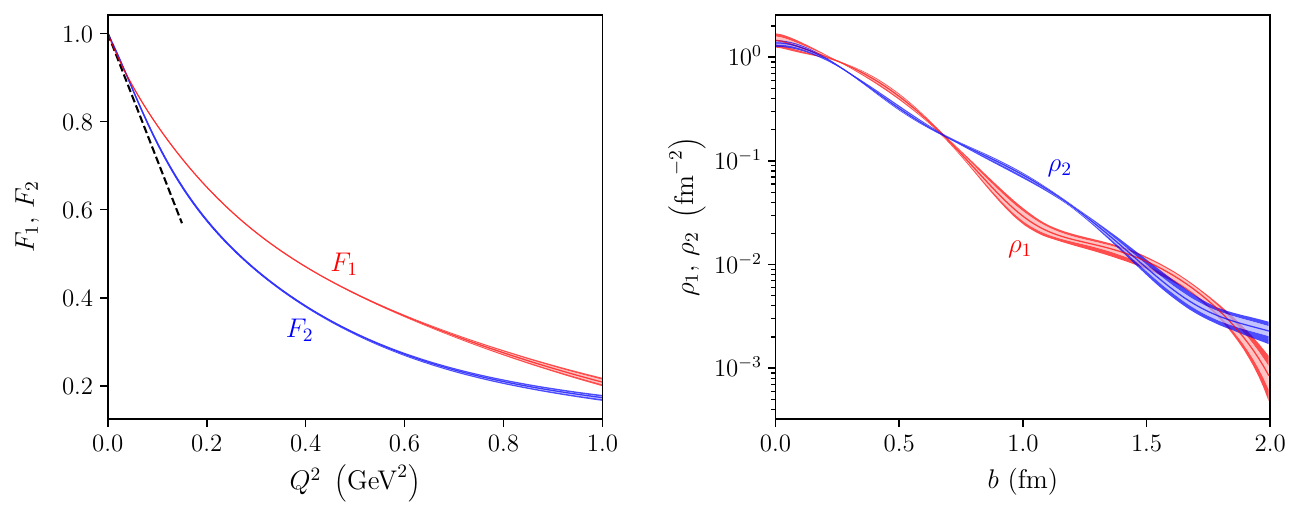}
\caption{\label{fig:fit_plots}Form factors and transverse densities extracted using our best model ($N = 5$, $\lambda = 0.02$). Left panel: form factors $F_1$ (red) and $F_2$ (blue) as functions of~$Q^2$. The black dashed line is a tangent to $F_1$ at $Q^2 = 0$ corresponding to the mean-square transverse charge radius $\langle b_1^2 \rangle = 11.49~\text{GeV}^{-2}$. Note that our extraction of $\langle b_1^2 \rangle$ is based on Eq.~(\ref{eq:b1_integral}) rather than Eq.~(\ref{eq:b1_derivative}). Right panel: transverse densities $\rho_1$ (red) and $\rho_2$ (blue) as functions of~$b$. In both panels, lighter inner bands indicate the 68\% statistical confidence intervals of the corresponding quantities, while darker outer bands show the 68\% statistical and systematic confidence intervals added in quadrature.}
\end{figure*}

\begin{figure*}
\vspace{\baselineskip}
\includegraphics[width=\textwidth]{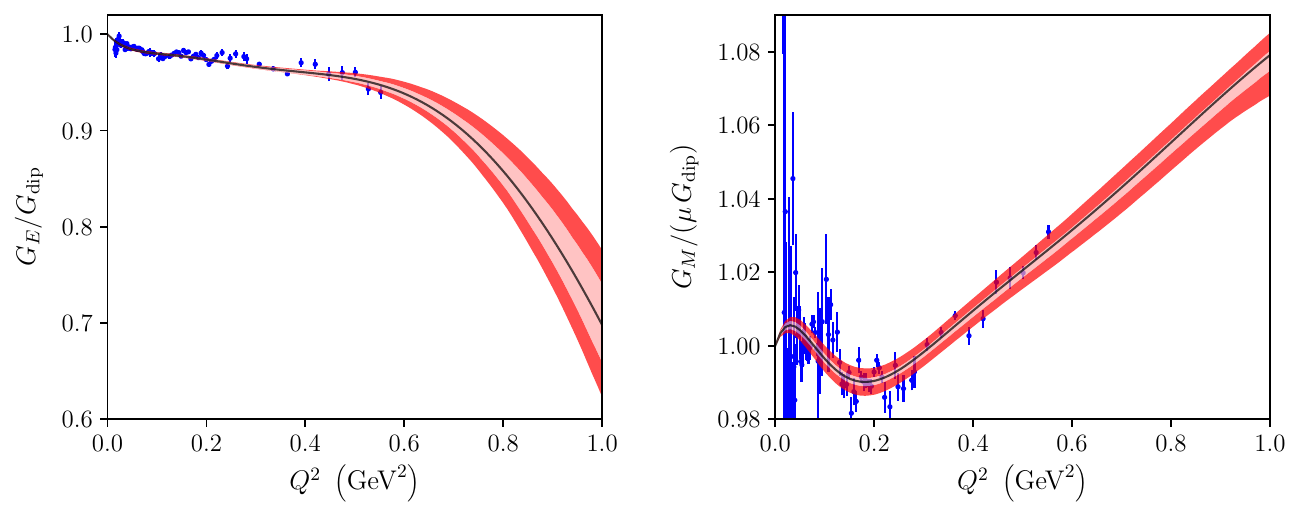}
\caption{\label{fig:ge_gm}Extracted electric (left panel) and magnetic (right panel) form factors as functions of~$Q^2$. We determine $G_E$ and $G_M$ from $F_1$ and $F_2$ using Eq.~(\ref{eq:GE_GM}) and scale them by the corresponding dipole form factors (see Appendix~\ref{app:cross_section}). The lighter inner bands around the black best-fit lines are the 68\% statistical confidence intervals, while the darker outer bands are the 68\% statistical and systematic confidence intervals added in quadrature. We compare our extraction with the values of $G_E$ and $G_M$ obtained in Ref.~\cite{Bernauer_PRC.90.015206} using the Rosenbluth separation technique (blue data points with error bars representing statistical uncertainties).}
\end{figure*}

The form factors and the transverse densities given by our main fit are shown in Fig.~\ref{fig:fit_plots} with 68\% confidence intervals. The point-to-point (statistical) uncertainties are determined by propagating the errors of the fit parameters taking into account the full covariance matrix. To estimate the systematic uncertainties, we follow the original analysis and refit our model using four modifications of the cross section data. These modifications correspond to the upper and lower bounds of (1)~the energy cut in the elastic tail and (2)~all other systematic effects linear in the scattering angle~\cite{Bernauer_PRC.90.015206}. We perform all fits with floating normalizations and use the largest deviation from the primary fit as an uncertainty estimate.

As an additional check, we also extract the form factors $G_E(Q^2)$ and $G_M(Q^2)$ and confirm that they agree with the values obtained in Ref.~\cite{Bernauer_PRC.90.015206} using the Rosenbluth separation technique (see Fig.~\ref{fig:ge_gm}). Note that the Rosenbluth results are model-independent but based only on a subset of the full A1 cross section data. For further details on the data analysis, the reader is referred to our \textsc{Python} code~\cite{GitHub}. The A1 Collaboration data that we use are publicly available as the Supplemental Material of Ref.~\cite{Bernauer_PRC.90.015206}.

Our final extraction of the proton charge radius from the full A1 data yields
\begin{equation}
r_E = 0.889(5)_{\text{stat}}(5)_{\text{syst}}(4)_{\text{model}}~\text{fm}, \label{eq:our_radius}
\end{equation}
where the model misspecification uncertainty of 0.004~fm is estimated based on the higher-order values of~$r_E$ listed in Table~\ref{tab:results}. Our radius is larger by 0.01~fm than the original result~(\ref{eq:A1_radius}), but both values are consistent given their uncertainties. Therefore, we confirm that the A1 data imply a large proton radius, although the possibility of unrecognized systematic errors can never be ruled out.

\section{Conclusion}

We have presented a novel method for extracting the proton charge radius from elastic scattering data that does not require determining the slope of $G_E$ at $Q^2 = 0$. The method is based on Eq.~(\ref{eq:r_E_new}) relating $r_E$ to the second moment of the transverse charge density~$\rho_1(b)$. This density is the two-dimensional Fourier transform of the Dirac form factor~$F_1(Q^2)$ and has a proper relativistic interpretation. As a consequence, $\rho_1(b)$ and $r_E$ can be determined by analyzing all available scattering data, not just those obtained at low~$Q^2$ values. Another novelty is the use of $F_1$ instead of the usual~$G_E$ to extract the proton radius. To facilitate the analysis, we have proposed reasonable parametrizations not only for the form factors $F_1(Q^2)$ and $F_2(Q^2)$, but also for the transverse densities $\rho_1(b)$ and $\rho_2(b)$.

We have applied our method to the extensive data obtained by the A1 Collaboration~\cite{Bernauer_PRL.105.242001, Bernauer_PRC.90.015206}. To find the right balance between underfitting and overfitting, we have used cross-validation and regularization---best practices from the field of statistical learning often overlooked in nuclear physics. Figure~\ref{fig:fit_plots} shows the form factors and the transverse densities that we have extracted. Our method has yielded the proton radius~(\ref{eq:our_radius}), which is consistent with the A1 value~(\ref{eq:A1_radius}) but larger by 0.01~fm. Therefore, our reanalysis has confirmed that the full A1 data lead to the proton charge radius that contradicts the muonic hydrogen results~\cite{Pohl_Nature.466.213, Antognini_Science.339.417}. This means that the discrepancy cannot be explained by issues with data fitting and extrapolation. Further progress can be achieved by combining our approach with a careful reanalysis of all available electron-proton scattering data. Finally, the method can be extended to better understand other properties of the proton such as its magnetic radius and higher-order density moments.

\begin{acknowledgments}
We thank Prof. J.~C.~Bernauer for his comments that helped to improve the manuscript.
\end{acknowledgments}

\appendix
\section{Choice of form factor parametrization}
\label{app:parametrization_choice}

Our choice of parametrizations for $\rho_1(b)$, $\rho_2(b)$, $F_1(Q^2)$, and $F_2(Q^2)$ was motivated by the following requirements.
\begin{enumerate}
\item We should be able to analytically calculate both the forward (\ref{eq:F1_transform}) and the inverse (\ref{eq:rho1_transform}) Fourier transforms. This requirement ensures that closed-form expressions exist for both the form factors and the transverse densities.

\item The form factors should have the correct static limit, $F_1(0) = F_2(0) = 1$, which corresponds to the following normalization for $\rho_1(b)$ and $\rho_2(b)$:
\begin{equation}
\int\limits_{0}^{\infty} b \rho_1(b) \, db = \int\limits_{0}^{\infty} b \rho_2(b) \, db = \frac{1}{2\pi}. \label{eq:rho_normalization}
\end{equation}

\item The form factors should have reasonable asymptotic behavior---for example, the large-$Q^2$ behavior predicted by the dimensional scaling laws~\cite{Brodsky_PRD.11.1309}: $F_1(Q^2) \propto Q^{-4}$ and $F_2(Q^2) \propto Q^{-6}$.

\item The parametrization should form a complete set of basis functions or, at least, should be flexible enough to approximate any real experimental data (with as few terms as possible).
\end{enumerate}
As shown in the main text, all four criteria are satisfied if we use the expansions (\ref{eq:rho1_expansion}) for $\rho_1(b)$ and (\ref{eq:rho2_expansion}) for $\rho_2(b)$, which correspond to the rational parametrizations (\ref{eq:F1_expansion}) for $F_1(Q^2)$ and (\ref{eq:F2_expansion}) for $F_2(Q^2)$. Moreover, because of the orthogonality condition~(\ref{eq:orthonormality}), there is a simple one-to-one correspondence~(\ref{eq:b1_moments}) between the expansion coefficients $\alpha_n$ and the even moments $\langle b_{1}^{2n} \rangle$ of the transverse charge density $\rho_1(b)$.

Although we find this choice particularly convenient and elegant, other parametrizations can also be used. Consider, for example, a Pad\'e approximant, which was first applied to form factors by Kelly~\cite{Kelly_PRC.70.068202}. In the case of $F_1(Q^2)$, the Pad\'e approximant of order $[K / (K+2)]$ should be used, which takes the form
\begin{equation}
F_1(Q^2) = \frac{1 + \sum_{n=1}^{K} a_n Q^{2n}}{1 + \sum_{n=1}^{K+2} b_n Q^{2n}}, \label{eq:Kelly_F1}
\end{equation}
where $a_n$ and $b_n$ are $2(K + 1)$ free parameters. A similar expression can be written for the Pauli form factor by using the $[K / (K+3)]$ approximant. The parametrization~(\ref{eq:Kelly_F1}) satisfies most of the above criteria, but does not allow for an analytical expression for $\rho_1(b)$. Moreover, depending on specific values of the coefficients $b_n$, the function~(\ref{eq:Kelly_F1}) can have poles for $Q^2 \ge 0$ (at the points where the denominator is zero). Such poles are unphysical and make the Fourier integral~(\ref{eq:rho1_transform}) divergent. Nevertheless, the parametrization~(\ref{eq:Kelly_F1}) can still be used, provided that the poles are avoided for $Q^2 \ge 0$ (for example, by choosing $b_n \ge 0$) and that a closed-form expression for $\rho_1(b)$ is not required.

Note that our form factor parametrization is simply a special case of the one suggested by Kelly. Indeed, Eq.~(\ref{eq:F1_expansion}) for $F_1(Q^2)$ can be rewritten with a common denominator as
\begin{equation}
F_1(Q^2) = \frac{1 + \sum_{n=1}^{2N} \tilde \alpha_n \bigl(Q^2 / \Lambda^2\bigr)^n}{\bigl(1 + Q^2 / \Lambda^2\bigr)^{2N + 2}}, \label{eq:our_F1}
\end{equation}
where $\tilde \alpha_n$ are $2N$ linear combinations of the $N$ expansion coefficients $\alpha_i$. Equations (\ref{eq:Kelly_F1}) and (\ref{eq:our_F1}) coincide after choosing $K = 2N$, $a_n = \tilde \alpha_n \Lambda^{-2n}$, and
\begin{equation}
b_n = \binom{2N+2}{n} \, \Lambda^{-2n},
\end{equation}
where the parentheses denote the binomial coefficient. An advantage of our parametrization~(\ref{eq:our_F1}) is that it guarantees the absence of poles in $F_1(Q^2)$ for $Q^2 \ge 0$.

It is important to realize that many form factor parametrizations used in previous extractions of the proton radius do not correspond to bounded transverse densities. We argue that such parametrizations should be rejected as unphysical. For example, if one chooses a polynomial parametrization for $F_1(Q^2)$ then the Fourier integral (\ref{eq:rho1_transform}) diverges. The same happens if one uses polynomial fits for $G_E(Q^2)$ and $G_M(Q^2)$. A suitable form factor parametrization should correspond to bounded transverse densities that can be normalized as in Eq.~(\ref{eq:rho_normalization}).

Finally, we note that for any suitable parametrization the value of $\langle b_1^2 \rangle$ determined as a second moment~(\ref{eq:b1_integral}) of the transverse charge density will agree with the value (\ref{eq:b1_derivative}) obtained through the derivative of $F_1(Q^2)$ at $Q^2 = 0$. This does not mean, however, that Eqs.~(\ref{eq:b1_integral}) and (\ref{eq:b1_derivative}) are equivalent. First of all, many form factor parametrizations do not correspond to bounded transverse densities and should therefore be rejected, regardless of their slope at $Q^2 = 0$. Second, the integral formula~(\ref{eq:b1_integral}) applies to both even and odd moments of the transverse charge density, while the derivative formula is limited to even moments only:
\begin{equation}
\langle b_1^{2n} \rangle = (-4)^n n! \, F_{1}^{(n)}(0), \label{eq:derivative_formula}
\end{equation}
where $F_{1}^{(n)}(0)$ denotes the $n$th derivative of $F_1$ with respect to~$Q^2$ evaluated at $Q^2 = 0$. Therefore, we argue that Eq.~(\ref{eq:b1_integral}) for the $n$th moment $\langle b_1^n \rangle$ is more general than Eq.~(\ref{eq:derivative_formula}) and its special case, Eq.~(\ref{eq:b1_derivative}).

\section{Elastic scattering cross section}
\label{app:cross_section}

We use the beam energy, $E$, and the negative four-momentum transfer squared, $Q^2$, as two independent kinematic variables. The electron scattering angle, $\theta$, can be determined from $E$ and~$Q^2$ as
\begin{equation}
\theta = \arccos{\left[1 - \frac{M Q^2}{E \left(2M E - Q^2\right)}\right]}.
\end{equation}
Also useful are the dimensionless kinematic variables $\tau$ and $\varepsilon$, defined as
\begin{gather}
\tau = \frac{Q^2}{4M^2}, \\
\varepsilon = \left[1 + 2(1 + \tau) \tan^2{\frac{\theta}{2}}\right]^{-1}.
\end{gather}

The differential cross section for unpolarized elastic electron-proton scattering is given by the Rosenbluth formula
\begin{equation}
\frac{d \sigma_0}{d \Omega} = \frac{\sigma_{\text{red}}}{\varepsilon (1 + \tau)} \frac{d \sigma_{\text{Mott}}}{d \Omega}, \label{eq:Rosenbluth}
\end{equation}
where
\begin{align}
\sigma_{\text{red}} &= \varepsilon \, G_E^2(Q^2) + \tau \, G_M^2(Q^2) \nonumber \\
&= \varepsilon \left[F_1(Q^2) - \tau \kappa F_2(Q^2)\right]^2 \nonumber \\
&+ \tau \left[F_1(Q^2) + \kappa F_2(Q^2)\right]^2 \label{eq:sigma_red}
\end{align}
is the so-called reduced cross section and $d \sigma_{\text{Mott}} / d \Omega$ is the Mott cross section describing the scattering of electrons on spinless point charged particles. The Sachs form factors $G_E$ and $G_M$ are often approximated as
\begin{gather}
G_E(Q^2) \approx G_{\text{dip}}(Q^2), \label{eq:G_E_dip} \\
G_M(Q^2) \approx \mu \, G_{\text{dip}}(Q^2), \label{eq:G_M_dip}
\end{gather}
where
\begin{equation}
G_{\text{dip}}(Q^2) = \left(1 + \frac{Q^2}{0.71~\text{GeV}^2}\right)^{-2}
\end{equation}
is the standard dipole form factor and $\mu = 1 + \kappa$ is the magnetic moment of the proton. The corresponding reduced cross section is
\begin{equation}
\sigma_{\text{dip}} = \left(\varepsilon + \mu^2 \tau\right) G_{\text{dip}}^2(Q^2). \label{eq:sigma_dip}
\end{equation}

\section{Fit consistency and $Q^2$ sensitivity}
\label{app:consistency}

Here we study the consistency and $Q^2$ sensitivity of our proton radius extraction approach assuming a variety of different underlying form factor models.

As a first study, similar to the approach in Refs.~\cite{Kraus_PRC.90.045206, Yan_PRC.98.025204}, we preform a series of Monte Carlo simulations by generating cross section pseudodata at the A1 experimental kinematic points and point-to-point uncertainties assuming a specific form factor model and that the errors are Gaussian. We do not account for systematic and normalization uncertainties in these simulated experiments. By truncating the pseudodata above a range of maximum $Q^2$ values and redoing the full fit, we can explore how the high-$Q^2$ data affect our extracted radius variance and bias for different orders~$N$ of our parametrization. We study six diverse form factor models: dipole (\ref{eq:G_E_dip})--(\ref{eq:G_M_dip}), inverse polynomial (Arrington-2004~\cite{Arrington_PRC.69.022201}), low-order Pad\'e approximant (Kelly~\cite{Kelly_PRC.70.068202}), continued fraction expansion (Arrington-2007~\cite{Arrington_PRC.76.035201}), high-order Pad\'e approximant (Venkat~\cite{Venkat_PRC.83.015203}), and polynomial (Bernauer~\cite{Bernauer_PRC.90.015206, Bernauer_thesis}). We note that the two Pad\'e approximant models, Kelly~\cite{Kelly_PRC.70.068202} and Venkat~\cite{Venkat_PRC.83.015203}, have fit parameter values that result in a bounded transverse charge density and are thus considered physically plausible in our approach. The Bernauer polynomial fit~\cite{Bernauer_PRC.90.015206, Bernauer_thesis} was obtained in the original analysis of the A1 Collaboration data.

\begin{figure*}[t!]
\includegraphics[width=\textwidth]{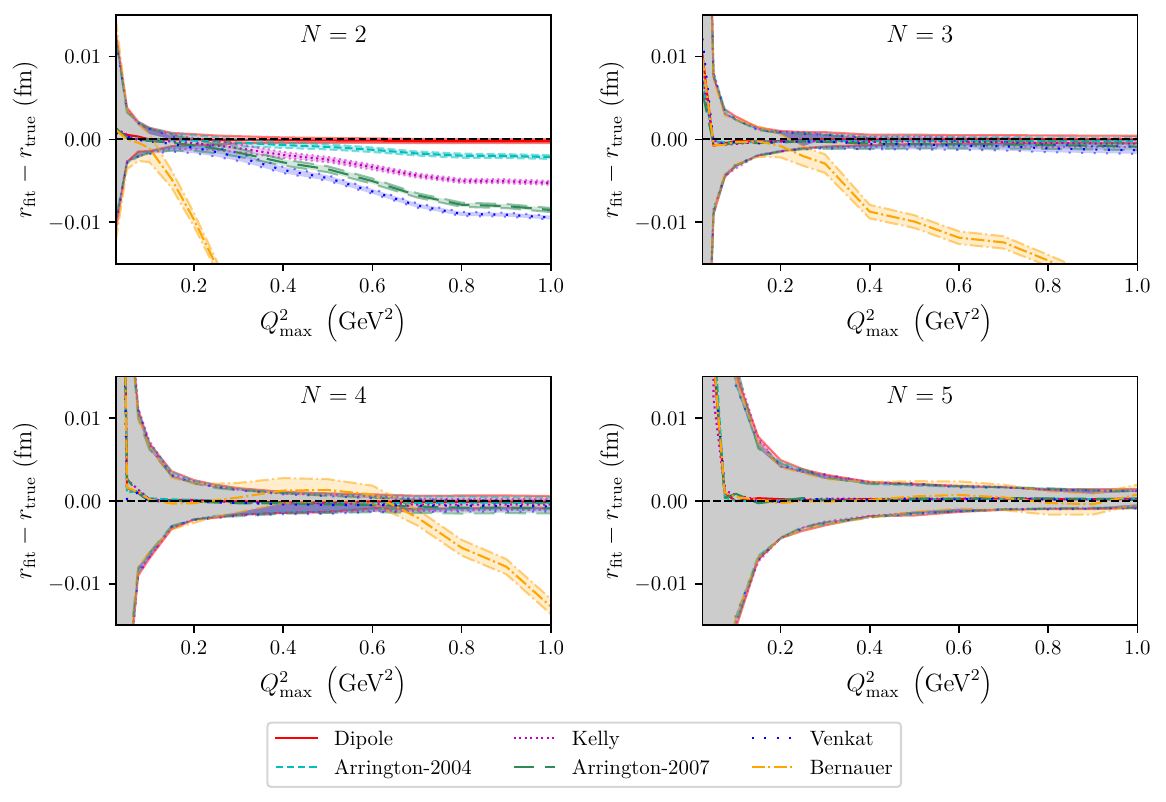}
\caption{\label{fig:consistency_study}The difference between the extracted and the true values of the proton radius as a function of $Q_{\text{max}}^2$ cutoff for orders $N = 2$, 3, 4, and 5 of our parametrization. For this study, we use pseudodata generated with six different form factor models: dipole (\ref{eq:G_E_dip})--(\ref{eq:G_M_dip}), Arrington-2004~\cite{Arrington_PRC.69.022201}, Kelly~\cite{Kelly_PRC.70.068202}, Arrington-2007~\cite{Arrington_PRC.76.035201}, Venkat~\cite{Venkat_PRC.83.015203}, and Bernauer~\cite{Bernauer_PRC.90.015206, Bernauer_thesis}. Lines and shaded bands represent the mean values and the standard deviation intervals, respectively, for the quantity $r_{\text{fit}} - r_{\text{true}}$ obtained in repeated experiments.}
\end{figure*}

The results of this study are shown in Fig.~\ref{fig:consistency_study} for orders $N = 2$, 3, 4, and 5 of our parametrization, with no additional regularization used. Each of the four panels in Fig.~\ref{fig:consistency_study} corresponds to specific~$N$ and shows the difference, $r_{\text{fit}} - r_{\text{true}}$, between the extracted and the true underlying values of the proton radius as a function of the $Q_{\text{max}}^2$ cutoff in the pseudodata. This figure is a good illustration of the bias-variance trade-off, where ``bias'' is given by the deviation of the mean difference $r_{\text{fit}} - r_{\text{true}}$ from zero and ``variance'' is shown by the width of the standard deviation band. As expected, the simpler underlying models are well fit by the lower-order parametrizations, while the more complex models require higher-order parametrizations. The model complexity can be roughly estimated by the number of free parameters used (this number ranges from 1 for the dipole model to 20 for Bernauer's polynomial~\cite{Bernauer_PRC.90.015206, Bernauer_thesis}). For the simplest---dipole---model, the order $N = 2$ is sufficient. The most complex model---the polynomial fit obtained in the original A1 analysis---demands for $N = 5$. Note that the $N = 5$ fit performs well for all of the considered models.

Although instructive, tests on pseudodata cannot replace cross-validation for choosing the optimal order~$N$ because existing models do not necessarily capture the full complexity of the proton form factors. Interestingly, the underfitted (i.e., biased) extractions in Fig.~\ref{fig:consistency_study} tend to produce smaller values of the proton radius. This observation agrees with the conclusion made in Ref.~\cite{Kraus_PRC.90.045206} for polynomial fits. Perhaps this effect may be responsible for the smaller values of the proton radius obtained in some of the earlier reanalyses of the A1 dataset.

Figure~\ref{fig:consistency_study} also shows how the statistical precision of the extracted radius improves as higher-$Q^2$ data are added, while bias does not increase as long as $N$ is sufficiently large to accommodate the complexity of the underlying model. It might be tempting to limit bias by fitting less complex functions to the lowest-$Q^2$ data, but this comes at the price of increased variance. Higher-$Q^2$ data help to better constrain the fit and to achieve higher precision. Therefore, the best approach is to analyze all available data and use cross-validation for finding the right balance between bias and variance.

\begin{figure*}
\includegraphics[width=\textwidth]{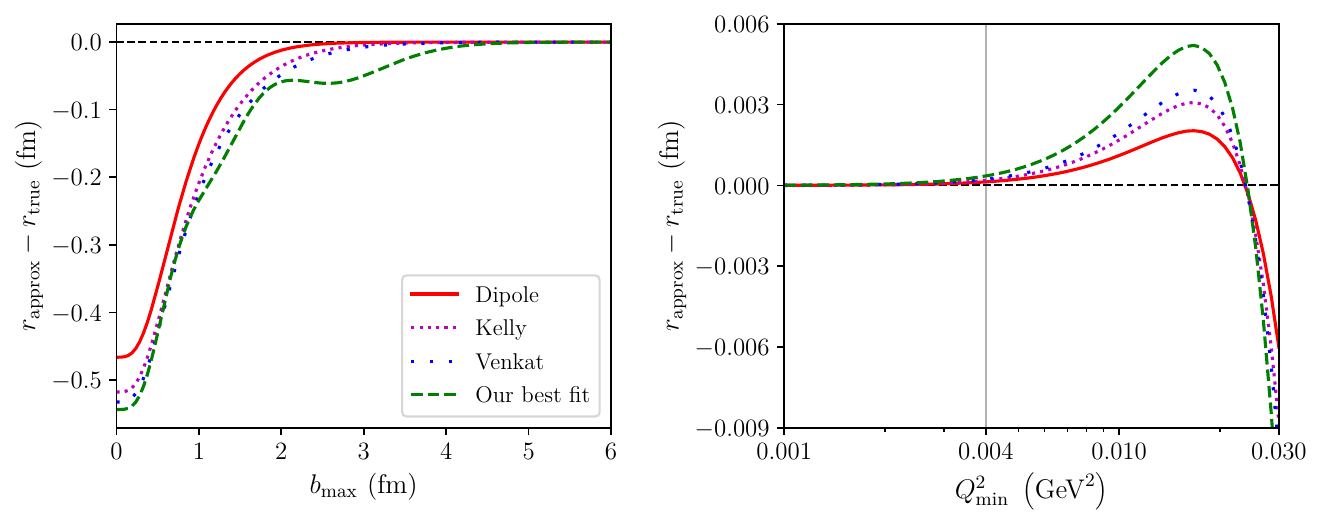}
\caption{\label{fig:Q2_min}Sensitivity of the extracted radius to the data at high $b$ (left panel) and low $Q^2$ (right panel). For this study, we use four underlying form factor models: dipole (\ref{eq:G_E_dip})--(\ref{eq:G_M_dip}), Kelly~\cite{Kelly_PRC.70.068202}, Venkat~\cite{Venkat_PRC.83.015203}, and our best fit (see Table~\ref{tab:coeffs}). The gray vertical line in the right panel at $0.004~\text{GeV}^2$ corresponds to the lower bound of the $Q^2$ range covered by the A1~experiment.}
\end{figure*}

To estimate the sensitivity of the extracted radius to low-$Q^2$ data, including below the reach of the A1 experiment, we use a different approach. We first note that there is an inverse relationship between the variables $Q$ and $b$: low-$Q^2$ behavior of the form factor~$F_1(Q^2)$ corresponds to the tail of the transverse charge density $\rho_1(b)$ at large~$b$. Since the transverse size of the proton is of the order of $\sqrt{\langle b_1^2 \rangle} \approx 0.67~\text{fm}$ and the density $\rho_1(b)$ rapidly decays with increasing~$b$, we expect that the region $b \gg 1~\text{fm}$ does not contribute significantly to the second moment~(\ref{eq:b1_integral}). To show this quantitatively, we use the four different form factor models that correspond to bounded transverse charge densities: dipole (\ref{eq:G_E_dip})--(\ref{eq:G_M_dip}), Kelly~\cite{Kelly_PRC.70.068202}, Venkat~\cite{Venkat_PRC.83.015203}, and our best fit (see Table~\ref{tab:coeffs}). We obtain the corresponding densities $\rho_1(b)$ by calculating the Fourier integral~(\ref{eq:rho1_transform}) numerically. We then evaluate the second moment of $\rho_1(b)$,
\begin{equation}
\langle b_1^2 \rangle_{\text{approx}} = 2\pi \int\limits_0^{b_{\text{max}}} b^3 \rho_1(b) \, db, \label{eq:b2_approx}
\end{equation}
by varying the upper integration limit~$b_{\text{max}}$. We finally use Eq.~(\ref{eq:r_E_new}) to determine the corresponding proton radius, $r_{\text{approx}}$. Our results for the difference $r_{\text{approx}} - r_{\text{true}}$ are shown in the left panel of Fig.~\ref{fig:Q2_min} as a function of~$b_{\text{max}}$. It turns out that the upper integration limit of $b_{\text{max}} = 5~\text{fm}$ is sufficient for determining the proton radius with accuracy better than $0.001~\text{fm}$ for all of the considered form factor models. This result allows us to roughly estimate the lower bound of the required $Q^2$ range as $Q_{\text{min}}^2 \approx (\hbar c / b_{\text{max}})^2 \approx 0.002~\text{GeV}^2$.

To obtain a more accurate estimate of the sensitivity in $Q^2$ space, we assume that $F_1$ is measured down to $Q^2_\text{min}$ and then linearly interpolated to $F_1(0) = 1$, giving us:
\begin{equation*}
F_{1}^{\text{approx}}(Q^2)
=
\begin{cases}
1 - \left[1 - F_1(Q_{\text{min}}^2)\right] \frac{Q^2}{Q_{\text{min}}^2}, & Q^2 < Q_{\text{min}}^2, \\
F_1(Q^2), & \text{otherwise}.
\end{cases}
\end{equation*}
We then find the corresponding transverse charge density by numerically calculating the integral~(\ref{eq:rho1_transform}) from $Q = 0$ to $Q = \infty$. Note that we do not simply use $Q_{\text{min}}$ as the lower integration limit in Eq.~(\ref{eq:rho1_transform}) because this would make $\rho_1(b)$ oscillating, as expected from the properties of Fourier transform. After that, we obtain the second moment of $\rho_1(b)$ by using Eq.~(\ref{eq:b2_approx}) with the upper integration limit $b_{\text{max}} = 6~\text{fm}$. We finally substitute $\langle b_1^2 \rangle_{\text{approx}}$ into Eq.~(\ref{eq:r_E_new}) and determine the corresponding radius $r_{\text{approx}}$. Our results are shown in the right panel of Fig.~\ref{fig:Q2_min} as a function of $Q_{\text{min}}^2$. We can conclude from the figure that the value $Q_{\text{min}}^{2} = 0.004~\text{GeV}^2$, which corresponds to the lower bound of the $Q^2$ range covered by the A1 Collaboration, is sufficient for determining the proton radius with accuracy better than 0.001~fm.

To summarize, we have shown in this Appendix that the $Q^2$ range covered by the A1 experiment---from 0.004 to $1~\text{GeV}^2$---is sufficient to achieve a low-bias, low-vari\-ance determination of the proton radius. We emphasize that, in addition to reaching the small $Q^2$ values, it is equally important to have a sufficiently wide $Q^2$ coverage.

\section{Model dependence and model misspecification error}
\label{app:model_dependence}

As the true functional form of the proton form factors is unknown, the extraction of the proton radius from scattering data is inherently model dependent. We refer to the bias in the radius estimate due to any mismatch between the model representation and the true form factors as the model misspecification error. While the uncertainty from this error source cannot be precisely quantified, it is essential to study and deliberately minimize its effect. As we showed in Appendix~\ref{app:consistency}, our parametrization yields negligible model misspecification error for a variety of form factor models, though this is not guaranteed to be the case for the true proton form factors. Our approach contains the following components to reduce the amount of model misspecification error in our extraction.
\begin{enumerate}
\item Our parametrizations are highly flexible and able to approximate any true underlying transverse densities and form factors given an appropriate order $N$.

\item As discussed in Appendix~\ref{app:parametrization_choice}, our parametrizations are constrained to fit known physics of the proton form factors. In particular, our fit functions for $F_1(Q^2)$ and $F_2(Q^2)$ have the correct static limit, $F_1(0) = F_2(0) = 1$, and the asymptotic behavior expected from the dimensional scaling laws~\cite{Brodsky_PRD.11.1309}: $F_1(Q^2) \propto Q^{-4}$ and $F_2(Q^2) \propto Q^{-6}$. Importantly, our approach ensures that the extracted transverse charge density $\rho_1(b)$ is bounded, which constrains the set of plausible fit functions without adding model misspecification bias.

\item We use data-driven model selection (i.e., cross-validation) over the entire available $Q^2$ range of data to find the best balance between underfitting and overfitting. Note that the optimal model complexity cannot be determined before the data are collected since the complexity of the true form factors is unknown. For example, models selected based on pseudodata generated assuming simple underlying parametrizations may be too biased, as shown in Appendix~\ref{app:consistency}. Cross-validation allows us to estimate the total error, including model misspecification error, of our model on unseen data. By selecting our model complexity (through $N$ and $\lambda$) to minimize the cross-validation error, we also minimize the model misspecification error in our fit.
\end{enumerate}

In addition to taking these three critical steps to minimize model misspecification error, we also estimate the magnitude of the uncertainty due to it in our extraction by using an ensemble of regularized higher-order fits. For each higher order of $N$, we independently optimize $\lambda$ to minimize the cross-validation error on the full dataset. The spread in this ensemble of models, shown in Table~\ref{tab:results}, allows us to estimate the model misspecification uncertainty of our proton radius extraction (including the effect of the floating normalization parameters required by the A1 dataset) as 0.004~fm.


\onecolumngrid
\newpage

\renewcommand{\theequation}{S\arabic{equation}}

\mdfdefinestyle{MyFrame}{%
    linewidth=1pt,
    innertopmargin=4pt,
    innerbottommargin=4pt,
    innerrightmargin=4pt,
    innerleftmargin=4pt}

\phantomsection
\label{sec:supplement}

\begin{center}
\textbf{\large Supplemental Material for\\ ``Transverse charge density and the radius of the proton''}
\end{center}

Polynomials~$P_{n}^{(\nu)}(x)$ are defined by the orthonormality condition given in Eq.~(\ref{eq:orthonormality}) of the main text. The first six terms of~$P_{n}^{(1)}(x)$ are
\begin{flalign}
P_{0}^{(1)}(x) &= 1,& \\
P_{1}^{(1)}(x) &= \frac{x - 2}{2\sqrt{2}},& \\
P_{2}^{(1)}(x) &= \frac{x^2 - 15 \, x + 18}{6 \sqrt{26}},& \\
P_{3}^{(1)}(x) &= \frac{13 \, x^3 - 636 \, x^2 + 5328 \, x - 4896}{144 \sqrt{4303}},& \\
P_{4}^{(1)}(x) &= \frac{331 \, x^4 - \num{37620} \, x^3 + \num{997200} \, x^2 - \num{6105600} \, x + \num{4708800}}{\num{224640} \, \sqrt{1986}},& \\
P_{5}^{(1)}(x) &= \frac{676 \, x^5 - \num{148095} \, x^4 + \num{8964900} \, x^3 - \num{169866000} \, x^2 + \num{844128000} \, x - \num{572702400}}{\num{1123200} \, \sqrt{\num{1375726}}}.&
\end{flalign}
The first six polynomials~$P_{n}^{(2)}(x)$ are
\begin{flalign}
P_{0}^{(2)}(x) &= 1,& \\
P_{1}^{(2)}(x) &= \frac{x - 3}{\sqrt{15}},& \\
P_{2}^{(2)}(x) &= \frac{5 \, x^2 - 96 \, x + 168}{24 \sqrt{110}},& \\
P_{3}^{(2)}(x) &= \frac{11 \, x^3 - 645 \, x^2 + 6840 \, x - 9000}{360 \sqrt{1738}},& \\
P_{4}^{(2)}(x) &= \frac{79 \, x^4 - \num{10368} \, x^3 + \num{327420} \, x^2 - \num{2517120} \, x + \num{2743200}}{1440 \sqrt{\num{11938638}}},& \\
P_{5}^{(2)}(x) &= \frac{\num{25187} \, x^5 - \num{6219045} \, x^4 + \num{433086780} \, x^3 - \num{9730620900} \, x^2 + \num{60331975200} \, x - \num{57256264800}}{\num{151200} \, \sqrt{\num{541950039098}}}.&
\end{flalign}
The rational functions~$A_{n}(y)$ are defined by Eq.~(\ref{eq:A_N}) in the main text. The first six terms are
\begin{flalign}
A_{0}(y) &= \frac{1}{\left(1 + y\right)^2},& \\
A_{1}(y) &= -\frac{y \, (y + 4)}{\sqrt{2} \left(1 + y\right)^4},& \\
A_{2}(y) &= \frac{y^2 (3 \, y^2 + 22 \, y + 39)}{\sqrt{26} \left(1 + y\right)^6},& \\
A_{3}(y) &= -\frac{y^3 (34 \, y^3 + 352 \, y^2 + 1187 \, y + 1324)}{\sqrt{4303} \left(1 + y\right)^8},& \\
A_{4}(y) &= \frac{y^4 (1635 \, y^4 + \num{21560} \, y^3 + \num{104885} \, y^2 + \num{225774} \, y + \num{182520})}{78 \sqrt{1986} \left(1 + y\right)^{10}},& \\
A_{5}(y) &= -\frac{y^5 (\num{13257} \, y^5 + \num{210730} \, y^4 + \num{1324320} \, y^3 + \num{4152432} \, y^2 + \num{6528139} \, y + \num{4127178})}{26 \, \sqrt{\num{1375726}} \left(1 + y\right)^{12}}.&
\end{flalign}
The rational functions~$B_{n}(y)$ are defined by Eq.~(\ref{eq:B_N}) in the main text. The first six terms are
\begin{flalign}
B_{0}(y) &= \frac{1}{\left(1 + y\right)^3},& \\
B_{1}(y) &= -\frac{\sqrt{3} \, y \, (y + 5)}{\sqrt{5} \left(1 + y\right)^5},& \\
B_{2}(y) &= \frac{y^2 (7 \, y^2 + 64 \, y + 132)}{\sqrt{110} \left(1 + y\right)^7},& \\
B_{3}(y) &= -\frac{y^3 (25 \, y^3 + 321 \, y^2 + 1260 \, y + 1580)}{\sqrt{1738} \left(1 + y\right)^9},& \\
B_{4}(y) &= \frac{\sqrt{3} \, y^4 (635 \, y^4 + \num{10324} \, y^3 + \num{58410} \, y^2 + \num{141476} \, y + \num{125935})}{\sqrt{\num{3979546}} \left(1 + y\right)^{11}},& \\
B_{5}(y) &= -\frac{y^5 (\num{378679} \, y^5 + \num{7377979} \, y^4 + \num{53840262} \, y^3 + \num{189977062} \, y^2 + \num{329168959} \, y + \num{225929067})}{\sqrt{\num{541950039098}} \left(1 + y\right)^{13}}.&
\end{flalign}
The first ten terms of $P_{n}^{(1)}$, $P_{n}^{(2)}$, $A_{n}$, and $B_{n}$ can be calculated using the following \textsc{Mathematica} code:
\begin{mdframed}[style=MyFrame]
\begin{Verbatim}[commandchars=\\\{\}, codes={\catcode`$=3}]
\$Assumptions\:=\:y\:>\:0;
w[$\nu$_]\::=\:2*x^($\nu$/2)*BesselK[$\nu$,\,2*Sqrt[x]]/Gamma[$\nu$+1]
P1\:=\:Together[Orthogonalize[x^Range[0,\,9],\,Integrate[#1*#2*w[1],\,\{x,\,0,\,Infinity\}]\,&]]
P2\:=\:Together[Orthogonalize[x^Range[0,\,9],\,Integrate[#1*#2*w[2],\,\{x,\,0,\,Infinity\}]\,&]]
A\:=\:Together[Integrate[#*w[1]*BesselJ[0,\,2*Sqrt[x*y]],\,\{x,\,0,\,Infinity\}]\,&\,/@\,P1]
B\:=\:Together[Integrate[#*w[2]*BesselJ[0,\,2*Sqrt[x*y]],\,\{x,\,0,\,Infinity\}]\,&\,/@\,P2]
\end{Verbatim}
\end{mdframed}

\end{document}